\documentclass[conference]{IEEEtran}
\IEEEoverridecommandlockouts
\usepackage{array}
\usepackage{booktabs}
\usepackage{multirow}
\usepackage{bm}
\usepackage{cite}
\usepackage{amsmath,amssymb,amsfonts}
\usepackage{algorithmic}
\usepackage{graphicx}
\usepackage{textcomp}
\usepackage{xcolor}
\usepackage[normalem]{ulem}
\usepackage{hyperref}
\def\BibTeX{{\rm B\kern-.05em{\sc i\kern-.025em b}\kern-.08em
    T\kern-.1667em\lower.7ex\hbox{E}\kern-.125emX}}

\usepackage{algorithm}
\usepackage{algorithmic}

\usepackage{bm}

\usepackage{authblk}

\usepackage{ifthen}
\newboolean{authnotes}
\setboolean{authnotes}{True} 
\ifthenelse{\boolean{authnotes}}
{
\newcommand{\cl}[1]{\footnote{\color{orange}{\bf CL: #1}}}
\newcommand{\jz}[1]{\footnote{\color{purple}{\bf JZ: #1}}}
\newcommand{\lw}[1]{\footnote{\color{blue}{\bf LW: #1}}}

\usepackage{draftwatermark}
\SetWatermarkLightness{0.95}
\SetWatermarkScale{1}
}
{
\newcommand{\cl}[1]{}
\newcommand{\jz}[1]{}
\newcommand{\lw}[1]{}

}

\begin{document}

\title{Progressive Defense Against Adversarial Attacks for Deep Learning as a Service in Internet of Things}

\author[1]{Ling Wang}
\author[1]{Cheng Zhang}
\author[1]{Zejian Luo}
\author[2]{Chenguang Liu}
\author[3]{Jie Liu}
\author[4]{Xi Zheng}
\author[5]{Athanasios Vasilakos}
\affil[1]{Department of Computer Science and Technology, Harbin Insititute of Technology }
\affil[2]{The University of Texas at Austin}
\affil[3]{Department of Computer Science and Technology, Harbin Insititute of Technology(ShenZhen) }
\affil[4]{Department of Computing, Macquaire University}
\affil[5]{School of Electrical and Data Engineering, University of Technology Sydney}
\affil[ ] { }
\affil[1]{\textit{wangling@hit.edu.cn, \{18S103130, 20S003073\}@stu.hit.edu.cn}}

\affil[2]{\textit{liuchg@utexas.edu}}
\affil[3]{\textit{jieliu@hit.edu.cn}}
\affil[4]{\textit{james.zheng@mq.edu.au}}
\affil[5]{\textit{Athanasios.Vasilakos@uts.edu.au}}


\maketitle
\pagestyle{plain}

\begin{abstract}
Nowadays, Deep Learning as a service can be deployed in Internet of Things (IoT) to provide smart services and sensor data processing. However, recent research has revealed that  some Deep Neural Networks (DNN) can be easily misled by adding relatively small but adversarial perturbations to the input (e.g., pixel mutation in input images). One challenge in defending DNN against these attacks is to efficiently identifying and filtering out the adversarial pixels.
The state-of-the-art defense strategies with good robustness often require additional model training for specific attacks. To reduce the computational cost without loss of generality, 
we present a defense strategy called a progressive defense against adversarial attacks (PDAAA) for efficiently and effectively filtering out the adversarial pixel mutations, which could mislead the neural network towards erroneous outputs, without {\it a-priori} knowledge about the attack type.
We evaluated our progressive defense strategy against various attack methods on two well-known datasets. The result shows it outperforms the state-of-the-art while reducing the cost of model training by 50\% on average.

\end{abstract}



\section{Introduction}\label{sec:intro}


Deep learning involves building and training a neural network, a machine learning model inspired by the human brain. Once a neural network is trained on a dataset, it can be used for a variety of recognition tasks. However, developing deep learning models is a painstakingly iterative and experimental process often requiring hundreds, even thousands of training runs that need very large amount of computing power to find the right combination of neural network configurations and hyperparameters.  To simplify this neural network building process and making it possible even for professionals without deep coding experience to do it, Deep Learning as a Service (DLaaS) has become the current trend for using DL techniques.
DLaaS can be deployed in many area of IoT like self-driving cars,  smart warehouse, smart surgery where DNN plays very important roles\cite{li2020unsupervised,yue2019privacy,zhang2020achieving}.
The progression from data to intelligence usually consists of two phases. In the training phase, the network learns how to do a job through intensive computation. The trained network will then be used in the inference phase, where it infers things about the real world input based on the knowledge it acquired in the training. For example, image recognition networks are commonly used to detect objects from camera input and make predictions on what are presented in the image (e.g., a pedestrian, two bicyles)~\cite{deng2020analysis}.

However, recent works have exposed a significant vulnerability in DNN where small perturbations to the input can  ``fool'' the model and cause it to produce wrong outputs\cite{szegedy2013intriguing} 
One example commonly used in the autonomous driving field is the adversarial patch. In this method, attaching small stickers to the traffic sign could confuse the recognition network to make wrong predictions. In a recent study, self-driving models predicting steering angles (regression models instead of classification models which are usually studied in the literature) are found vulnerable to adversarial attacks as well~\cite{deng2020analysis}.
Among existing defense approaches, the method that filtering out pixel mutations  can achieve defense success rate without changing network structure\cite{akhtar2018defense,xu2017feature2,meng2017magnet,gebhart2017adversary}. However, one constraint that can greatly impact the feasibility of applying existing defense methods is the device capability, especially for the edge devices in IoT scenarios~\cite{zhang2020fenghuolun}.
Our work tries to fill this gap thereby enhance the security aspect of the deep learning based real-time IoT services through a low-cost yet practical defense method. One of the key insights of our approach is the proportion of pixel mutations that need to be filtered out in an image to achieve effective defense is usually small. As such, our method focuses on filtering out small number of pixels based on differential evolution to reduce the operations applied on the input sample thereby minimizing the computation cost.

Our  key contributions are summarized as follows: 

\begin{enumerate}
\item We define the sensitive points property of the neural network input as the partial perturbations that are likely to disrupt the prediction accuracy. Based on this definition, we formulate a novel problem to effectively search for the sensitive points in neural networks.
\item We present a systematic approach, namely progressive defense against adversarial attacks (PDAAA), which can obtain sensitive points for adversarial pixel mutations and filter them in the following defense. 
\item Empirical experiments show that our approach outperforms the state-of-the-art while reducing computation time by 50\% on average.
\end{enumerate}

The paper is organized as follows. In the section \ref{sec:background} and \ref{sec:relatedwork}, we introduce the background and related work of  our approach. The proposed problem and defense approach are presented in the section  \ref{sec:problem_def} and \ref{sec:strategy}. Section \ref{sec:expsetting} and \ref{sec:expresult} provide our experimental setup and result analysis  followed by the conclusion section \ref{sec:conclusion} .




\section{Background}\label{sec:background}

Deep learning is a class of of machine learning methods based on artificial neural networks.
Deep learning can realize  image classification by learning the characteristics of data representation.  The excellent feature extraction ability of deep learning makes it a great success in many fields such as computer vision and IoT \cite{wang2020intelligent, zhang2020fenghuolun, zhang2020democratically, xu2020tt}.
Many complicated problems solved by deep learning have reached or even surpassed the level of human beings~\cite{lu2020efficient,wang2020leveraging1}. However, the research has revealed that deep learning can be vulnerable just like other technologies\cite{szegedy2013intriguing, huang2017adversarial}. Given the deployment scale of IoT applications, how to secure our devices and services against adversarial attacks (e.g., input perturbation) becomes an imminent issue~\cite{deng2020analysis}.

An adversarial example is an input crafted by an adversary aiming to cause incorrect outputs from a target classifier. Since ground truth, at least for image classification tasks, is based on human perception which is hard to model or test, research in adversarial examples typically defines an adversarial example as a misclassified sample generated by perturbing a correctly-classified sample by some limited amount\cite{goodfellow2014explaining,kurakin2016adversarial,madry2017towards}. Small perturbation is a fundamental premise for adversarial
examples. The magnitude of perturbation can be measured by $p$-norm distance. 
\begin{equation}
L_p = || X - X^{adv}||_p = (\sum \limits_{i=1}^n |X_i - X_i^{adv}|^p)^{\frac 1 p}  \label{eq2-1}
\end{equation}

here, $X$ represents an input sample, and  $X^{adv}$ represents an adversarial sample. $L_0$, $L_2$,and $L_{\infty}$ are three commonly $L_p$ used metrics. $L_0$ counts
the number of pixels changed in the adversarial examples.
$L_2$ measures the Euclidean distance between the adversarial example and the original sample. $L_{\infty}$ denotes the maximum change for all pixels in adversarial examples.




\section{Related work}\label{sec:relatedwork}
Training and inferencing deep neural networks, known as deep learning, is currently highly complex and
computationally intensive, especially for deployment in IoT, edge computing and mobile clouds.  Instead of being mired with infrastructure and cluster management problems, users would like to focus on training a model in the easiest way possible that satisfies both their cost and performance objectives. This is where the opportunity of deep learning as a service lies. IBM has provided DLaaS platform from their cloud where the infrastructure is shared across these workloads while providing a common API-based
access. Challenging security issues has to be faced, such as security and privacy issues on systems and algorithms of DLaaS. Among these issues, deep learning security on defense against adversarial attacks becomes the most important one. 
\subsection{Adversarial Attack Methods}\label{sec:adversarial_attack_methods}

\subsubsection{Global Attack Methods}

Global attack means that the attack method adds disturbance to the whole input sample. Some pixels have a larger disturbance range, while others have smaller disturbance ranges.


 
Goodfellow et al.\cite{goodfellow2014explaining} have found that the previous explanation of the existence of confrontation samples is attributed to the nonlinearity and overfitting of the network, while they have made the opposite conclusion through experiments that the existence of confrontation samples is due to the linear characteristics of the network. Based on this conclusion, they have proposed a simple and fast method to generate confrontation samples, which can be used for confrontation training. It is called as a fast gradient symbol method (FGSM). 
 
 



BIM attack method was proposed by kurakin et al.\cite{kurakin2016adversarial} in 2016. BIM method is an improvement result of FGSM. The method is to modify the pixel value one step at a time by iteration and prune the generated pixel value to ensure that it is within the given range of the original image. Compared with FGSM, this method is more imperceptible. 




PGD \cite{madry2017towards} is an iterative attack, which can be regarded as an iterative version of FGSM. We know that BIM is also an iterative version of FGSM. The difference between the two is that PGD has more iterations and adds randomization, which makes the attack effect of PGD better than FGSM, and can generate more imperceptible adversarial disturbance than FGSM. The attack principle is basically the same as that of BIM, that is, the pixel value is modified one step at a time by iteration, and the generated pixel value is pruned to ensure that it is within the given range of the original image $X$.

It can be found that the above methods involve the gradient information of loss function relative to the input sample. The calculation of the gradient is global, and its calculation will affect the value of each pixel. 
\subsubsection{Local Attack methods}
Local attack means that the attack method adds disturbance to one or more pixels in the whole input sample. The number or percentage of pixels is often a super parameter, which can be specified artificially. 

In this kind of attack method, the evolutionary algorithm is often used to find out the key components which have a great influence on the output of the deep learning model, and then small disturbance is added to the key components to generate countermeasures. The advantage of this attack method is that it does not need to know the parameters of the deep learning model, and it can successfully attack some non-differentiable network models. The main methods of Adversarial attack based on the evolutionary algorithm are as follows: One-Pixel Attack \cite{su2019one} and LocSearchAdv \cite{narodytska2017simple}.

\subsection{Defense Methods}
The defense methods can be roughly divided into three kinds: (1) training data approaches that modify the training data during the training phase or the input sample during the test phase \cite{moosavi2017universal,bhagoji2018enhancing,xu2017feature1,goodfellow2014explaining}; (2) network structure approaches that change the network structure, such as changing activation and loss functions\cite{papernot2016limitations,ross2017improving,lyu2015unified}; (3) disturbance filter approaches that filter out the disturbance in the sample before the test sample is passed into the original network \cite{akhtar2018defense,xu2017feature2,meng2017magnet,gebhart2017adversary}.

For the training data approach, defense methods mainly include adversarial training, data compression, and data randomization. Adversarial training refers to taking a large number of adversarial samples as training data, modifying their labels as correct values and then passing them into the neural network, so that the trained model can learn to resist adversarial samples. This method was first proposed by \cite{rumelhart1986learning}. However, it was pointed out that adversarial training can only make the model have good robustness for the adversarial samples in the training set, however, theoretically, there are infinitely adversarial samples \cite{moosavi2017universal}. Therefore, no matter how many adversarial samples are added, there still exist new adversarial attack samples that can deceive the network again. Since more adversarial samples should be generated for adversarial training, the computational requirements of adversarial training are very large. The data compression method refers to the adoption of a compression algorithm for the input sample, which makes the modification of pixel be obvious to the neural network as they are smaller after compression. The method was proposed to compress the input data by principal component analysis to obtain the robustness of adversarial training \cite{bhagoji2018enhancing}. However, it was shown that such compression would also damage the spatial structure of the image, thus often have a negative impact on the performance \cite{xu2017feature1}. The data randomization method refers to performing some randomization operations on the input sample, such as translation, rotation, clipping, scaling, and filling. Attacks were successfully defended for L-BFGS and FGSM by translating the input samples \cite{goodfellow2014explaining}. However, this method can only protect against some weak attacks and is ineffective against more powerful attacks.

The network structure approach is usually to change the loss function or activation function, which makes it tolerant of a small adversarial disturbance without changing the output result of the training model. The results show that this method has good robustness against attacks such as FGSM and JSMA \cite{papernot2016limitations, ross2017improving}. The robustness of the network was improved against L-BFGS and FGSM attacks by regularizing the loss function \cite{lyu2015unified}. However, each of these approaches almost doubles the training complexity of the network.

The disturbance filter approach is different from two others as this approach neither changes the original network nor directly modifies the data, but adds a model to handle the disturbance. Then, the output of the input sample is processed as the input of the original network. The method was proposed to add a defense network to correct the disturbed images by training the defense network to make the classification prediction of adversarial samples consistent with that of undisturbed samples \cite{akhtar2018defense}. The original network learns classification and the defense network learns correction against disturbance. This kind of defense methods can effectively correct the image without changing the original network and data. The shortcoming is that the perturbation filtering mechanism of the defense network is global filtering that leads to waste a lot of computational power. Different filtering methods have different effects on the input samples.

\section{problem definition} \label{sec:problem_def}
\begin{figure}[htbp]
\centerline{\includegraphics[width=0.30\textwidth]{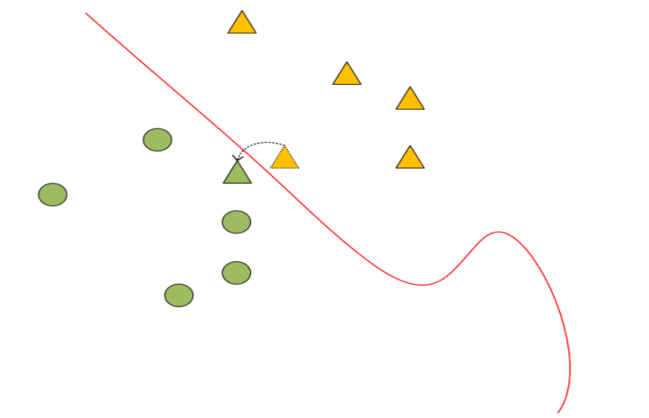}}
\caption{The decision boundary of classified samples}
\label{fig:4}
\end{figure}

A common idea for creating adversarial images is adding a tiny amount of well-tuned additive perturbation, which is expected to be imperceptible to human eyes, to a correctly classified natural image. Such modification can cause the classifier to label the modified image as a completely different class. The classification of an image is determined by all pixels, which can be regarded as a high-dimensional vector. It can be mapped to high-dimensional space. Every change of the pixel value will make the position of the pixel vary in the space. And because the decision boundary is trained, it is fixed and determined by weights and biases. If the pixel position of the image is very close to the decision boundary, then a slight change in the coordinate of the pixel may cause the new position of the pixel to cross the decision boundary as shown in Fig. \ref{fig:4}. Thus, the wrong classification result is achieved.

Due to the above reasons, some attacks consider limited scenarios for adversarial attacks to avoid modification seen by human eyes. For instance, only one pixel is attacked in the extremely limited scenarios \cite{su2019one}. Thus, defense methods to filter out perturbation of all pixels will waste large time and computation. If only attacked pixels are filter out, image can be efficiently protected. As the attacked pixels are usually close to the decision boundary, the key issue here is to be searched for these pixels and then filter them out with less time and computation. Therefore, there are two steps for defense stagey to be performed. Firstly, it is to find out these pixels that are selected as attacked target. And secondly, the pixels are filtered out for defense. At first, these pixels are defined as sensitive point, as followings:

Definition 1: \textbf{Sensitive point} is referred to the pixels in the input sample where the slight change in the value of the pixel will cause the neural network to have a wrong result for prediction. For example, a $32 \times 32$ size RGB color picture, which is a dog, changes the value of the pixel at one location  and finds that the prediction classification generated by transferring the changed image into the neural network f is a kitten. The pixel at the above location is a sensitive point.

It should be noted that sensitive point may not be one pixel. Sensitive point is a collection of pixels that can make the predicted value of neural network $f$ fluctuate greatly, and can have multiple pixels. The search for sensitive points can be formalized into an optimization problem with constraints, that is, to make the prediction confidence of neural network f as low as possible under the condition of changing as few pixels as possible. The input image is expanded into a vector, represented by $X$, each element in the vector represents a pixel, represented by $x_i$, then

\begin{equation}
X = (x_1, x_2, \cdots, x_n)  \label{eq1}
\end{equation}

Assume that there are changes in the original pixel value of the input image, then

\begin{equation}
s(X) = (s_1, s_2, \cdots, s_n)  \label{eq2}
\end{equation}

The problem is searching sensitive points for the neural network $f$ can be expressed as to solve the optimization $e(X)^*$.

\begin{equation}
\begin{split}
\mathop{minimize}  \limits_{e(X)^*} \ & f(X+s(X)) \\
\mathop{s.t.} \  ||  s & (X) ||_0  \leq d
\end{split}
\label{eq3}
\end{equation}

Where $d$ represents the number of sensitive points. For example, when $d=1$ means that changing the pixel of a point in the entire input image will cause a large change in the predicted value of the neural network $f$, then this point is the sensitive point.

\section{Proposed defense strategy} \label{sec:strategy}

\subsection{Search for sensitive points}

There are many optimization algorithms that are able to solve the problem \ref{eq3}. Compared with the gradient descent and greedy search algorithm, differential evolution algorithm is relatively less affected by the local minimum value of the heuristic algorithm. And the differential evolution algorithm does not require that other gradient information is optimized. Thus the objective function does not have to be differentiable or known for this problem. Because some neural networks are not differentiable or in many cases it is unrealistic to calculate their gradients, it can be used for a wider range of optimization problems than gradient-based methods. In addition, in this paper, the problem is defined with strict constraints, that is, the number of sensitive points much less compared to the number of input image pixels. Differential evolution algorithm only has to know the tag of neural network probability value, without the need of category, structure, parameters for the neural network.  Differential evolution algorithm is used to solve this optimization problem.

Differential evolution algorithm is a kind of evolutionary algorithm. Its main steps are basically the same as those of other evolutionary algorithms, including mutation, crossover, and selection. The differential evolution algorithm realizes individual variation through different strategies. The common difference strategy is to randomly select two different individuals in the population and merge the vectors with the individual to be mutated after the vector difference scaling, i.e.

\begin{equation}
\begin{split}
v_i(g+1)=x_{r_1}(g) & +\alpha \cdot (x_{r_2}(g)-x_{r_3}(g)) \\
i \neq r_1 & \neq  r_2 \neq r_3
\end{split}
\label{eq4}
\end{equation}

Where, $v_i(g+1)$ is the intermediate of variation, $r_1$, $r_2$ and $r_3$ are three random numbers, $x_i(g)$ represents the $i^{th}$ individual in the g generation population, and $\alpha$ is the scaling factor.This difference strategy is adopted. In order to ensure the effectiveness of the solution in the process of evolution, it is necessary to ensure that the "genes" in "chromosomes" meet the boundary conditions. If the boundary conditions cannot be met, the "genes" will be generated randomly to meet the conditions. Crossover operation refers to the crossover operation between individuals of the g generation population and its mutated intermediates. Due to the randomness of crossover, the probability should be introduced. The specific method of crossover operation is as follows

\begin{equation}
    u_{j,i}(g+1) = \begin{cases}
    1 \quad if \ rnd(0,1) \leq CR \ or \ j = j_{rnd} \\
    0 \quad otherwise
    \end{cases}
\label{eq5}
\end{equation}

Where, $CR$ is the crossover probability, $j_{rnd}$ is a random integer, and $v_{j,i}(g+1)$ is the element value of the $j^{th}$ position of the intermediate $v_i(g+1)$, whose value depends on $CR$.

The differential evolution algorithm uses the greedy algorithm to select individuals to enter the next generation of the population. The specific method is as follows:

\begin{equation}
    x_i(g+1) = \begin{cases}
    u_i(g+1)  & if \, f(u_i(g+1)) \leq f(x_i(g)) \\
    x_i(g)  & otherwise
    \end{cases}
\label{eq6}
\end{equation}

Here, $f(.)$ is the objective function that needs to be optimized.
In order to facilitate the positioning of pixels, we represent each pixel as
\begin{equation}
    \bm{x_i}=(x,y,r,g,b)
\label{eq7}
\end{equation}

Where $x$ and $y$ represent the coordinates of the pixel, $x$ is the distance of the pixel in the vertical direction relative to the top left corner of the origin, and $y$ is the distance of the pixel relative to the top left corner in the horizontal direction. $r$, $g$, and $b$ are the three primary colors of the pixel red, green, and blue, respectively, with values ranging from 0 to 255. Because the differential evolution algorithm requires the input value to be a one-dimensional vector, to facilitate the application of this algorithm, the input image is set as

\begin{equation}
    \bm{X=(x_1, x_2,\cdots, x_n)} 
\label{eq8}
\end{equation}

\begin{equation}
    X=(x_1, y_1, r_1, g_1, b_1, x_2, y_2, r_2, g_2, b_2, \cdots)
\label{eq9}
\end{equation}

$\bm{s(X)}$ is converted into the above format. First,  $\bm{s(X)}$ with number of m are randomly generated as the initial population, and then the following formula is used to generate m variants

\begin{equation}
\begin{split}
v_i(g+1)=x_{r_1}(g) & + \alpha \cdot (x_{r_2}(g)-x_{r_3}(g)) \\
i \neq r_1 & \neq  r_2 \neq r_3
\end{split}
\label{eq10}
\end{equation}

That is, each variant is randomly selected from three previous generations and then combined to form the next generation. The next generation of individuals is crossing over to form new ones with a certain probability. If the new individual $V_i$ can make the prediction probability value of neural network $f$ smaller than the previous generation individual $X_i$, then the previous generation individual will be eliminated. Repeating the above process several times until the corresponding sensitive points are found that make the prediction result of neural network $f$ wrong. The specific flow of the algorithm is given in Algorithm \ref{algo1} as follows.

\begin{algorithm}
\caption{Find Sensitive Points}

\begin{algorithmic}[1]

\REQUIRE Initialize the following parameters.

\begin{enumerate}[(1) ]
    \item the initial number of population: $popSize = 400$
    \item the scale parameter: $\alpha = 0.5$
    \item the probability of crossover: $CR = 0.8$
    \item the maximum number of iteration: $maxIter = 100$
    \item the number of sensitive points: $d = 10$
    \item the input image: $X$
\end{enumerate}

\ENSURE The coordinates of d sensitive points.

\STATE Initialize the first population: Randomly generate $popSize$ individuals $x_i$ 
\FOR{$g = 1 \to maxIter$}
    \FOR{$i = 1 \to popSize$}
        \STATE randomly choose three individuals to form a new candidate: 
        $v_i(g+1) \gets x_{r_1}(g) + \alpha \dots (x_{r_2}(g) - x_{r_3}(g))$
        \STATE Cross transformation with probability $CR$:
        \IF{$rnd(0, 1) \leq CR$ \textbf{or} $j = j_{rnd}$}
            \STATE $u_{j,i}(g+1) \gets v_{j,i}$
        \ELSE
            \STATE $u_{j,i}(g+1) \gets x_{j,i}(g)$
        \ENDIF
        \STATE each candidate solution compete with their corresponding father according to the index of the population and the winner survive for next iteration:
        \IF{$f(u_i(g+1)) \leq f(x_i(g))$}
            \STATE $x_i(g+1) \gets u_i(g+1)$
        \ELSE
            \STATE $x_i(g+1) \gets x_i(g)$
        \ENDIF
        \IF{the prediction label of neural network $f(\bm{x} + x_i(g+1))$ changes}
        \RETURN $x_i(g+1)$
        \ENDIF
    \ENDFOR
\ENDFOR

\end{algorithmic}
\label{algo1}
\end{algorithm}

\subsection{Filter out sensitive points}

The defense model we propose is shown in Figure 3, that is, the input samples will be sensitively filtered before being passed into the prediction model. And then the filtered input data will be passed into the prediction model for prediction. This type of defense can be reduced as much as possible, parameter changes in the prediction model. The trained model is easy to be transferred. The robustness is realized by adjusting the structure and parameters of the model before prediction.

\begin{figure}[htbp]
\centerline{\includegraphics[width=0.25\textwidth]{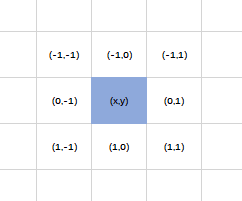}}
\caption{The relative coordinates of the eight points around the sensitive point (x,y) are shown}
\label{fig:6}
\end{figure}

Since the algorithm for finding sensitive points has been designed in the above, the next step is to filter out the sensitive points that can produce large fluctuations in the output of the prediction model for defense. The proposed filtering method is to take the average value of 8 pixels around each sensitive point as the filtered pixel value of the sensitive point to smooth the sensitive point. These 8 points are respectively located at the neighbor of the sensitive point, as shown in  Fig \ref{fig:6}. The left upper is set as the origin point of the coordinate of the pixel of the picture. Then,  the coordinates of the sensitive point is set as $(x,y)$ and the coordinates of the eight points around the sensitive point relative to this sensitive point are set as $(x-1,y-1),(x-1,y),(x-1,y+1),(x,y-1),(x,y+1),(x+1,y
-1),(x+1,y),(x+1,y+1)$.
The value at the filtered coordinate $(x, y)$ is defined as

\begin{equation}
\begin{split}
\overline{P}(x,y)=\frac 1 m \{ \sum \limits_{i=-1}^1 \sum \limits_{j=-1}^1 & P(x +i,y+j)-P(x,y) \} \\
\mathop{s.t.} \ 0 \leq x & + i < width \\
0 \leq y & + j < length 
\end{split}
\label{eq11}
\end{equation}

Where $width$ represents the width of the input image, $length$ represents the length of the input image. $P(x, y)$ represents RGB values at location $(x, y)$, $\overline{P}(x,y)$ represents c RGB values after filtered at coordinates of $(x, y)$. $m$ represents the number of pixels around  $(x, y)$ satisfying the constraints. It should be noted that the two constraints in the above equation must be specified because the number of pixels around each sensitive point is not fixed at 8. For example, when the coordinates of sensitive points are the first row, the last row, the first column or the last column, the number of pixels around will be less than 8.

For the convenience, we use the form of direction array to represent the coordinates of pixels around sensitive points, such as, dx = [-1,-1,-1,0,0,1,1,1], dy =[-1,0,1,-1,1,-1,0,1] respectively. dx, dy represent the relative transformation of coordinates of the X-axis and Y-axis. For instance, Location (1,1) is set at the last element of dx and dy. In other words, when looking for pixels around the center coordinate (x,y), the search order is  (-1,-1),(-1,0),(-1,1),(0,-1),(0,1),(1,-1),(1,0),(1,1). When finding coordinates beyond the length or width of the picture, it will skip this point and continue to search. For example, if the pixel on the upper left is searched that, x = x + dx[0], y = y + dy[0], and if the pixel on the lower right is searched that x = x + dx[7], y = y + dy[7]. If a valid pixel is not found, the number of pixels is m plus one. After traversing the direction array to find the surrounding pixels, the RGB value of these pixels are added, and then the average is assigned to the new RGB value of the sensitive points. The reason is that the sensitive point is the pixel values that have larger fluctuations in the output of the neural network prediction model. Since the sensitive points  have a large influence on the neural network prediction, they have the maximum difference with non-sensitive around pixels. Therefore, the average value of neighbor pixels is used to smooth the RGB value of sensitive points.

The process of constructing the defense strategy consists of two steps. The first step is to use the differential evolution algorithm to find the sensitive point coordinates, and the second step is to filter the sensitive points according to the sensitive point coordinates and the RGB pixel value of the input sample. The algorithm flow of filtering sensitive points is shown in Algorithm \ref{algo2}.

\begin{algorithm}

\caption{Filter Sensitive Points}

\begin{algorithmic}[1]

\REQUIRE Initialize the following parameters.

\begin{enumerate}
    \item the input image: $X$
    \item the coordinates of sensitives points: $SPC$
\end{enumerate}

\ENSURE The filtered image.

\STATE $dx \gets [-1, -1, -1, 0, 0, 1, 1, 1]$
\STATE $dy \gets [-1, 0, 1, -1, 1, -1, 0, 1]$

\FOR{$c = 0 \to SPC.size - 1$}
    \STATE $sum \gets 0$
    \STATE $num \gets 0$
    \FOR{$i = 0 \to 7$}
        \STATE $x \gets SPC[c].x + dx[i]$
        \STATE $y \gets SPC[c].y + dy[i]$
        \IF{$0 \leq x < X.width$ \textbf{and} $0 \leq y < X.length$}
            \STATE $sum \gets sum + X[x][y]$
            \STATE $num \gets num + 1$
        \ENDIF
    \ENDFOR
    \STATE $X[x][y] \gets \frac {sum} {num}$
\ENDFOR
\RETURN $X$

\end{algorithmic}
\label{algo2}
\end{algorithm}

\section{Experimental setting}\label{sec:expsetting}

\begin{table}[htbp]
  \centering
  \caption{Experiment Setting}
    \begin{tabular}[l]{ccc}
    \toprule
    & & Settings \\
    \midrule
    GPU   &       & GTX 1050 Ti \\
    Memory &       & 8GB \\
    Hard Disk &       & 500GB \\
    Operating System &       & Windows 10 \\
    Programming language &       & Python 3 \\
    Deep learning Frame &       & PyTorch, TensorFlow, Keras \\
    \bottomrule
    \end{tabular}%
  \label{tab1}%
\end{table}%

\subsection{Train Network Model}
Three ResNet, AlexNet and LeNet, are designed to test our defense strategy. The datasets are cifar-10\cite{krizhevsky2009learning} and MNIST\cite{lecun1998gradient}. The cifar-10 dataset is classified as aircraft, car, truck, boat, frog, bird, horse, cat, dog and deer. The size of each image is 32 × 32, and the RGB color channel is 3. The MNIST dataset category is 0 to 9.  Each image size is 28 × 28, and the black and white channel is 1. In other words, both data sets are 10 classification problems. The structures of LeNet and AlexNet models trained on cifar-10 dataset and MNIST are shown in table \ref{tab2} and table \ref{tab3}, respectively. ResNet50 is selected for ResNet model. There are three types of residual blocks. Their specific output size is shown in table \ref{tab4}.


\begin{table}[htbp]
  \centering
  \caption{ LeNet Network Structure}
    \begin{tabular}[l]{ccc}
    \toprule
    Type & & Structure \\
    \midrule
    Convolution Layer & & Kernel Size: (5,5); Number of kernels: 6 \\ 
Pooling Layer & & Kernel Size: (2,2); Maximum Pooling \\
Convolution Layer & &	Kernel Size: (5,5); Number of kernels: 60 \\ 
Pooling Layer & & Kernel Size: (2,2); Maximum Pooling \\
Convolution Layer & & Kernel Size: (5,5); Number of kernels: 1920 \\ 
Fully connection layer & &	pramenter: 121 $\times$ 84 \\
Output Layer & &	Number of neurons: 10; Softmax \\

    \bottomrule
    \end{tabular}%
  \label{tab2}%
\end{table}%

\begin{table}[htbp]
  \centering
  \caption{ AlexNet Network Structure}
   \begin{tabular}[l]{ccc}
    \toprule
    Type & & Structure \\
    \midrule
    Convolution Layer & & Kernel Size: (3,3); Number of kernels: 24 \\ 
    Pooling Layer & & Kernel Size: (2,2); Maximum Pooling \\ 
    Convolution Layer & & Kernel Size: (3,3); Number of kernels: 96 \\
    Pooling Layer & & Kernel Size: (2,2); Maximum Pooling \\ 
    Convolution Layer & & Kernel Size: (3,3); Number of kernels: 192 \\
    Convolution Layer & & Kernel Size: (3,3); Number of kernels: 192 \\
    Convolution Layer & & Kernel Size: (3,3); Number of kernels: 96 \\ 
    Pooling Layer & & Kernel Size: (3,3); Maximum Pooling \\
    Fully connection layer	& & pramenter: 98340 $\times$ 1024 \\
    Fully connection layer	& & pramenter: 1024 $\times$ 1024 \\
    Output Layer    && Number of neurons: 10; Softmax \\
    \bottomrule
    \end{tabular}%
  \label{tab3}%
\end{table}%

\begin{table}[htbp]
  \centering
  \caption{ResNet Residual Block Structure}
    \begin{tabular}[l]{ccc}
    \toprule
    Type & & Structure \\
    \midrule
    Residual Block 1 & & 32 $\times$ 32 $\times$ 16 \\
    Residual Block 2 & & 16 $\times$ 16 $\times$ 32 \\
    Residual Block 3 & & 8 $\times$ 8 $\times$ 64 \\
    \bottomrule
    \end{tabular}%
  \label{tab4}%
\end{table}%

After training the above three models, the classification accuracy of the verification set on cifar-10 and MNIST datasets is shown in table \ref{tab5}. We will use the model with this accuracy rate as the model used to generate adversarial samples and verify the defense effect.

\begin{table}[htbp]
  \centering
  \caption{the classification accuracy versus model}
    \begin{tabular}[l]{ccc}
    \toprule
    Model &  Cifar-10 & MNIST \\
    \midrule
    ResNet & 91.15\% & 97.25\% \\
    AlexNet & 78.46\% & 90.32\% \\
    LeNet & 72.69\% & 97.92\% \\ 

    \bottomrule
    \end{tabular}%
  \label{tab5}%
\end{table}%

It should be noted that if the real category of an original sample (without adding any disturbance) is inconsistent with the predicted category of the model, then the sample cannot be considered as an adversarial sample. This is because the model itself is not well trained, and the wrong classification is not caused by the attacker.The adversarial attack is to attack the predicted correct samples so that they are mispredicted. Therefore, AlexNet and lenet with a low accuracy rate on cifar-10 will not significantly affect our experiment results. The only impact is that there are fewer original samples that can be used to generate adversarial samples. However, this problem can be solved by generating multiple adversarial samples for the same original sample to ensure the total number of final adversarial samples.

\subsection{Generate adversarial samples}

In this paper, we use three kinds of attack methods: FGSM (fast gradient symbol method), BIM (basic iteration method), and PGD (projection gradient descent method). These three kinds of attack methods are commonly used in the existing defense strategy. We will use these three methods to generate adversarial samples, which can be used to test the defense effect in the next step.

When using FGSM attack method to generate adversarial samples, the variable $\varepsilon$ is used to control the amplitude of the disturbance. When the value of the variable is small, it is difficult to detect the difference between the generated adversarial sample and the original sample. The experiment shows that when $\varepsilon$ = 0.01, the proportion of adversarial samples generated by verification set is 1.30\%, Only 130 images in about 10000 verification sets are generated as adversarial samples. However, with the increasing of the value of $\varepsilon$, the proportion of images that are generated as adversarial samples will be significantly increased. For example, when $\varepsilon$ = 0.85, the number of images that are generated as adversarial samples in the verification set reaches 99.21\%. However, with the increase of the value of $\varepsilon$, the difference between the generated adversarial sample and the original sample will be particularly obvious. And even the human naked eye can no longer recognize the category of the image. We do not consider this kind of sample as an adversarial sample, because the sample not only deceives deep learning but also deceives the eyes of human beings. Therefore, we choose a value of $\varepsilon$, which can not only ensure the success ratio of the attack but also prevent the disturbance added to the original sample too obvious. The experiment shows that 0.55 is more appropriate. At this time, the proportion of adversarial samples generated by the verification set is 82.79\%. At this time, the adversarial samples generated on the cifar-10 data set are shown in Figure 4-10.

BIM is the iterative version of FGSM, and BIM contains the variable $\varepsilon$, and iterations. The main difference between BIM and FGSM is that BIM only updates a small step at each iteration, and then accumulates step by step. If deep learning can be cheated before the maximum number of iterations is reached, it will stop. If the maximum number of iterations has reached, deep learning still can not be cheated. Then the iteration will not be continued. This situation means that the original sample cannot be attacked by the BIM method. The number of iterations is a super parameter. Experiments show that more the iteration times may not have better performance. It is found that the original sample can not be attacked by BIM after more than ten iterations. If the iterations are continued, it will be useless and waste computing power. In the case of fixed iterations 50, the success rate of BIM attack cifar-10 dataset caused by different $\varepsilon$ values is shown in table \ref{tab4} - \ref{tab9}.

\begin{table}[htbp]
  \centering
  \caption{Attack success rate of BIM model under different $\varepsilon$}
    \begin{tabular}{cccc}
    \toprule
    \multirow{2}[4]{*}{Deep Learning Model} & \multicolumn{3}{c}{$\varepsilon$} \\
\cmidrule{2-4}    \multicolumn{1}{c}{} & 0.01  & 0.03  & 0.05 \\
    \midrule
    ResNet & 76.52\% & 93.15\% & 99.36\% \\
    AlexNet & 83.29\% & 95.02\% & 99.18\% \\
    LeNet & 90.63\% & 93.69\% & 99.73\% \\
    \bottomrule
    \end{tabular}%
  \label{tab6}%
\end{table}%

It can be found from table \ref{tab6} that the value of $\varepsilon$ is much smaller than that of the FGSM attack. This is because BIM is an iterative attack, and only a little disturbance is added each time. FGSM adds all disturbances at one time. Therefore, in order to make the one-time attack more likely, FGSM needs a larger value of $\varepsilon$ than BIM. In addition, it can be found that with the increasing of the value of $\varepsilon$, the attack success rate of BIM against the three models has been significantly improved. However, BIM is the same as FGSM that with the increase of the value of $\varepsilon$, the disturbance of the generated adversarial samples will be particularly obvious, and the original samples will be damaged greatly. Therefore, we take 0.03 as the value of $\varepsilon$, and the attack success rates of the three models are all greater than 90\%.

The PGD attack method is also the iterative version of FGSM. The difference between the PGD attack method and the BIM attack method is that PGD has more iterations.  The disturbance added in each iteration is smaller.  And the introduced randomness makes the generated adversarial samples more difficult to be detected by human eyes. In order to realize the PGD attack, we need to determine the disturbance amplitude $\varepsilon$ and iteration number iterations. Because PGD supports more rounds of iterations, the initial iteration number is 50 the final determined iteration number of BIM. And then the iteration times are increased continuously with the increment of 10 to find an appropriate iteration number for speeding up the generation of adversarial samples. The final determined value of BIM 0.03 is set as the initial value of the disturbance amplitude $\varepsilon$. The disturbance amplitude is continuously reduced with an increment of - 0.001 to find the optimal value. Therefore, the PGD attack method can not only ensure the quality of the generated adversarial samples but also improve the success rate of the attack and accelerate the generation of a large number of adversarial samples.

For each deep learning model, we generate 10000 adversarial samples according to the ratio of the training set to test set in data set under FGSM, BIM, and PGD attack methods to calculate the defense success rate. We know that the number of training sets / test sets of the cifar-10 data set is 50000 / 10000, the ratio of the training set to test set corresponding to MNIST data set is 60000 / 10000 The number of adversarial samples is shown in table \ref{tab7}.

As shown in Table VII, we randomly select 8333 images from the cifar-10 training set and 1667 images from the test set to generate adversarial samples. When the selected samples can not be generated as adversarial samples cannot, the solution is to select one randomly from the data set again as an alternative and repeat this operation until the adversarial samples can be generated. The operation of generating adversarial samples for the MNIST dataset is the same. Each attack method generates 10000 adversarial samples according to this method, which is used to verify the defense ability of defense methods against each attack method.

\begin{table}[htbp]
  \centering
  \caption{The number of adversarial samples generated by data set under three attack methods}
    \begin{tabular}[l]{cccc}
    \toprule
    Data set & Amount & Training set & Testing set \\
    \midrule
    Cifar-10 & 10000 & 8333 & 1667 \\
    MNIST & 10000 & 8571 & 1429 \\
    \bottomrule
    \end{tabular}%
  \label{tab7}%
\end{table}%

\section{Experiment Result}\label{sec:expresult}
In this section, we will verify the proposed defense strategy through experiments. Among them, there are three counter-attack methods: FGSM, BIM, and PGD. Three deep learning prediction models are ResNet, AlexNet, and LeNet. There are two data sets, cifar-10 and MNIST. 
The first step is to find the sensitive pixels in the adversarial sample, and the second step is to filter each sensitive point into a non-sensitive point. When the differential evolution algorithm is used to find sensitive points, the number of sensitive points is a super parameter. One input image may have one or more sensitive points. The number of sensitive points has to be set carefully because some key pixels determine the output category of the model. In addition, the more sensitive points are set, the more calculations will be made. Therefore, here the number of sensitive points D is 1, 10, 50, and 100, respectively. The defense effect of our designed defense method is verified on cifar-10 and MNIST data sets through experiments.

\subsection{Experiment result analysis on cifar-10}

The adversarial samples corresponding to the cifar-10 dataset have been generated in the previous section. Next, we only need to find and filter sensitive points for each adversarial sample, and then input the deep learning model to make predictions to judge whether the prediction classification is the real classification of the sample. If it is correct, it indicates that the defense against the adversarial sample is successful, otherwise, the defense fails. Because the operation of initial population generation and cross mutation involves random values when using differential evolution algorithm to find sensitive points, we calculate the defense success rate of the defense model as follows: independently calculate the defense success rate for three times, and take the average value as the final result. Table \ref{tab8} - \ref{tab11} show the success rate of defense method against attack when d = 1, d = 10, d = 50, d = 100, respectively.

\begin{table}[htbp]
  \centering
  \caption{Defense success rate when d=1}
    \begin{tabular}{cccc}
    \toprule
    \multirow{2}[4]{*}{Model} & \multicolumn{3}{c}{Adversarial attack methods} \\
\cmidrule{2-4}    \multicolumn{1}{c}{} & FGSM & BIM & PGD \\
    \midrule
    ResNet & 5.43\% & 0.26\% & 0.23\% \\
    AlexNet & 5.62\% & 0.73\% & 0.95\% \\
    LeNet & 5.94\% & 1.77\% & 1.42\% \\
    \bottomrule
    \end{tabular}%
  \label{tab8}%
\end{table}%

It can be found from Table \ref{tab8} that when the number of sensitive points is 1, the defense effect of the defense method on the three models is not high. When FGSM is used to attack the RESNET model, the defense method can only defend 5.43\% of the confrontation samples. When BIM and PGD are used to attack RESNET, the defense ability will be lower. This is mainly because in the 32 * 32 1024 pixels, only when the sensitive points found by the differential evolution algorithm are in the same position as the actual sensitive points in the anti-sample.  The probability itself is very low. Through comparison, it can be found that the defense ability of the defense method on LeNet and AlexNet is better than that of RESNET. This is because the structure of AlexNet and LeNet models is relatively simple. Compared with ResNet, it is more vulnerable to attack. The generated sample disturbance is larger, and it is easier to find sensitive points. In other words, the defense has worse performance for a more complex deep learning model and   attack method with less disturbance.

\begin{table}[htbp]
  \centering
  \caption{Defense success rate when d=10}
    \begin{tabular}{cccc}
    \toprule
    \multirow{2}[4]{*}{Model} & \multicolumn{3}{c}{Adversarial attack methods} \\
\cmidrule{2-4}    \multicolumn{1}{c}{} & FGSM & BIM & PGD \\
    \midrule
    ResNet & 52.23\% & 2.18\% & 1.69\% \\
    AlexNet & 52.89\% & 3.24\% & 2.98\% \\
    LeNet & 53.66\%	& 5.29\% & 3.41\% \\
    \bottomrule
    \end{tabular}%
  \label{tab9}%
\end{table}%

Table \ref{tab9} shows the defense success rate of different attack methods on three models when the sensitive points are specified as 10. It is shown that when d=10, the defense success rate will be improved than that of d=1, and the defense FGSM attack method is the most obvious. The same model has a much worse defense capability than BIM when facing the attack of BIM and PGD. The results show that the more sensitive points are found and filtered, the probability of successful defense will increase. This is because, in a certain range, with the increase of the number of specified sensitive points D, the more likely that the sensitive points found by the evolutionary algorithm are the same as the actual sensitive points against the sample.

\begin{table}[htbp]
  \centering
  \caption{Defense success rate when d=50}
    \begin{tabular}{cccc}
    \toprule
    \multirow{2}[4]{*}{Model} & \multicolumn{3}{c}{Adversarial attack methods} \\
\cmidrule{2-4}    \multicolumn{1}{c}{} & FGSM & BIM & PGD \\
    \midrule
    ResNet & 70.11\% & 17.18\% & 14.45\% \\
    AlexNet & 71.39\% & 20.34\% & 18.32\% \\
    LeNet & 73.08\% & 20.53\% & 21.16\% \\
    \bottomrule
    \end{tabular}%
  \label{tab10}%
\end{table}%

It can be found from table \ref{tab10} that the defense success rate of the three models for the FGSM attack is more than 70\%, which is much higher than that of BIM and PGD. The main reason is that the disturbance produced by the latter two attack methods is so small that it is difficult to find the actual sensitive point of the sample species. In addition, it can be seen from the table that when three different deep learning models face the same attack, the defense success rates of our defense methods are not very different. It indicates that the defense strategies we designed can be applied to various deep learning models.

\begin{table}[htbp]
  \centering
  \caption{Defense success rate when d=100}
    \begin{tabular}{cccc}
    \toprule
    \multirow{2}[4]{*}{Model} & \multicolumn{3}{c}{Adversarial attack methods} \\
\cmidrule{2-4}    \multicolumn{1}{c}{} & FGSM & BIM & PGD \\
    \midrule
    ResNet & 84.26\% & 32.25\% & 21.33\% \\
    AlexNet & 83.57\% & 36.68\% & 21.87\% \\
    LeNet & 85.13\% & 38.97\% & 23.16\% \\
    \bottomrule
    \end{tabular}%
  \label{tab11}%
\end{table}%

It can be found from Table \ref{tab11} that when the FGSM method is used to attack the RESNET model, the defense method can defend 84.26\% of the adversarial samples, and the number of designated sensitive points is only 100. It can be seen from the above table that the defense success rate of the three models in the face of the FGSM attack is very high. The success rate of defense against BIM attack is about 35\%, and that of PGD attack is more than 20\%. This shows that the idea of only modifying some pixels in the counter sample is feasible. At the same time, it shows that the defense method we designed can be widely used in different attack methods and deep learning models, and can deal with the attacks of various adversarial attack methods.

\begin{figure}[htbp]
\centerline{\includegraphics[width=0.45\textwidth]{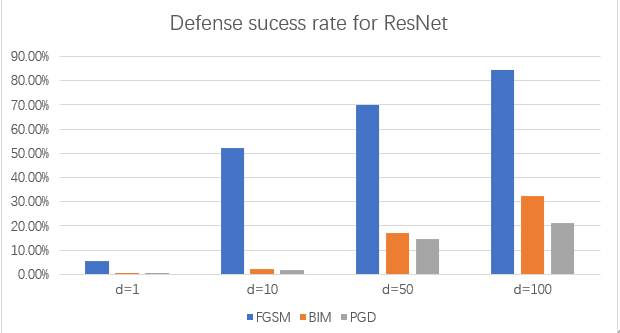}}
\caption{Defense success rate for ResNet by different attack methods}
\label{fig5}
\end{figure}

It can be seen from Figure 6 that the defense success rate on the RESNET model will increase with the increase of sensitive points D. The defense success rate of our designed defense method against FGSM, BIM, and PGD on RESNET model shows a decreasing trend. Among them, the defense ability of FGSM is improved most obviously. When d = 100, it can defend more than 80\% of the adversarial samples. The defense ability against the other two attack methods will also increase with the increase of sensitive points, but the growth rate is not very high. This is due to the small disturbance produced by the latter two attack methods.

\begin{figure}[htbp]
\centerline{\includegraphics[width=0.45\textwidth]{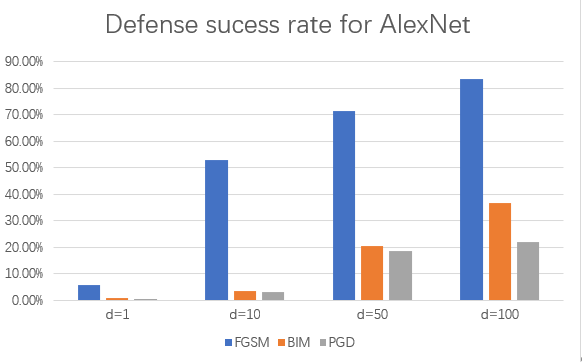}}
\caption{Defense success rate for AlexNet by different attack methods}
\label{fig6}
\end{figure}

From Figure 7, it can be found that the defense effect of our defense method on AlexNet is basically consistent with that of the ResNet model. The difference is that when the same sensitive point D is specified, the performance of the defense method on AlexNet is slightly better than that on ResNet.

\begin{figure}[htbp]
\centerline{\includegraphics[width=0.45\textwidth]{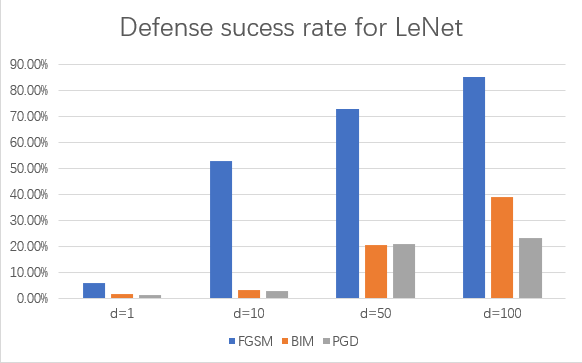}}
\caption{Defense success rate for LeNet by different attack methods}
\label{fig7}
\end{figure}

From Figure 6, Figure 7, and Figure 8, it is shown that the defense model designed by us can resist the attack of the counter samples generated on ResNet, AlexNet, and LeNet. The defense effect on LeNet is the best, followed by AlexNet and ResNet. But generally speaking, the performance of defense methods in the three models will not be very different. This shows that the defense method is not very sensitive to the differences in model structure parameters, and our proposed defense has generality.

\subsection{Experiment result analysis on MNIST}

Because each image in the MNIST dataset is black and white, it is different from that in the cifar-10 dataset. We select a random image from each category in the MNIST dataset as an example, as shown in Figure 9. It is necessary to verify the defense effect of our proposed defense method on the MNIST dataset. The adversarial samples corresponding to the MNIST dataset have been generated before. Next, we only need to perform defense operations on the adversarial samples, and then calculate the defense success rate. Its defense success rate is calculated in the same way as the cifar-10 dataset. Table \ref{tab12} - {tab15} show the success rate of defense method against attack when d = 1, d = 10, d = 50, d = 100, respectively.

\begin{figure}[htbp]
\centerline{\includegraphics[width=0.25\textwidth]{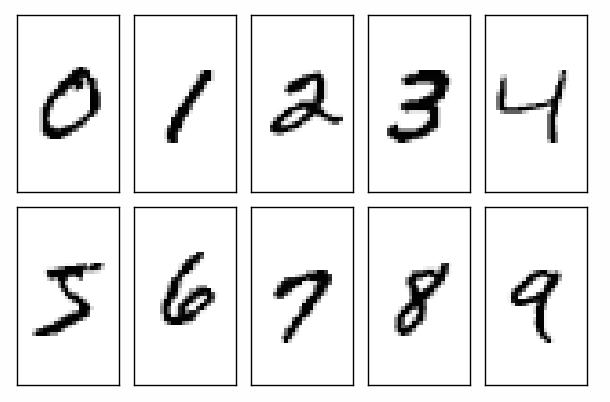}}
\caption{Mnist data set}
\label{fig8}
\end{figure}

\begin{table}[htbp]
  \centering
  \caption{Defense success rate when d=100}
    \begin{tabular}{cccc}
    \toprule
    \multirow{2}[4]{*}{Model} & \multicolumn{3}{c}{Adversarial attack methods} \\
\cmidrule{2-4}    \multicolumn{1}{c}{} & FGSM & BIM & PGD \\
    \midrule
    ResNet & 5.53\% & 1.17\% & 0.87\% \\
    AlexNet & 5.94\% & 1.24\% & 1.09\% \\
    LeNet & 6.35\% & 2.31\% & 1.65\% \\
    \bottomrule
    \end{tabular}%
  \label{tab12}%
\end{table}%

\begin{table}[htbp]
  \centering
  \caption{Defense success rate when d=100}
    \begin{tabular}{cccc}
    \toprule
    \multirow{2}[4]{*}{Model} & \multicolumn{3}{c}{Adversarial attack methods} \\
\cmidrule{2-4}    \multicolumn{1}{c}{} & FGSM & BIM & PGD \\
    \midrule
    ResNet & 54.41\% & 5.17\% & 5.20\% \\
    AlexNet & 54.64\% & 6.83\% & 5.54\% \\
    LeNet & 65.93\% & 5.89\% & 6.14\% \\
    \bottomrule
    \end{tabular}%
  \label{tab13}%
\end{table}%

\begin{table}[htbp]
  \centering
  \caption{Defense success rate when d=100}
    \begin{tabular}{cccc}
    \toprule
    \multirow{2}[4]{*}{Model} & \multicolumn{3}{c}{Adversarial attack methods} \\
\cmidrule{2-4}    \multicolumn{1}{c}{} & FGSM & BIM & PGD \\
    \midrule
    ResNet & 71.06\% & 20.57\% & 18.65\% \\
    AlexNet & 79.20\% & 23.25\% & 21.66\% \\
    LeNet & 80.37\% & 22.92\% & 23.84\% \\ 
    \bottomrule
    \end{tabular}%
  \label{tab14}%
\end{table}%

\begin{table}[htbp]
  \centering
  \caption{Defense success rate when d=100}
    \begin{tabular}{cccc}
    \toprule
    \multirow{2}[4]{*}{Model} & \multicolumn{3}{c}{Adversarial attack methods} \\
\cmidrule{2-4}    \multicolumn{1}{c}{} & FGSM & BIM & PGD \\
    \midrule
    ResNet & 80.06\% & 36.25\% & 31.27\% \\
    AlexNet & 87.27\% & 39.74\% & 34.10\% \\
    LeNet & 86.18\% & 40.61\% & 38.91\% \\
    \bottomrule
    \end{tabular}%
  \label{tab15}%
\end{table}%

From table \ref{tab11} - \ref{tab14} it can be found that the defense effect will be more obvious with the increase of the value of super parameter sensitive point D on the MNIST data set. When d = 100, the defense effect of the defense method on FGSM is the best. The three models are all above 80\%, and the best can reach 87.27\%. From the above table, we can see that the defense method we designed can play a defensive role against different attack methods on the basis of a variety of deep learning models. By comparing the results with the cifar-10 data set, it can be found that when the attack method is the same as the deep learning model, the defense success rate on the MNIST data set with simple structure is higher than that on the complex structure the cifar-10 data set. At the same time, it also shows that the defense method we designed can play a defense effect on different data sets, and has certain generality to the data sets.


\subsection{Comparison with existing methods}

For a black-box attack, the attacker has no knowledge about the target classifier or network, including the network structure and parameters of the classifier. It should be noted that Adversarial-PGD \cite{tramer2017ensemble}, PixelDefend \cite{wang2019direct}, Adversarial-Dual-defense \cite{yuan2020adversarial} have provided results for FSM and PGD attack on the cifar-10 dataset. The target classifier and detector network are based on the ResNet network. We compare our method with these three state-of-the-art methods under FGS and PGD attacks on the cifar-10 dataset as shown in table \ref{tab16}. From the table, we can see that our method has a better defense success rate than the existing method for the FGS attack. For the FGS attack, we improve upon the state-of-the-art method Adversarial-Dual-Network by 2.45\%. And we have the least computation time. The PGD attack is very powerful. It can totally fail the classifier with a resulting accuracy of 0\% if no defense is applied. For the PGD attack, we have the least defense success rate as d is set as 100 with the least computation time. As PGD is a multiple-order attack process, our method will improve the performance by selecting more pixels as sensitive points with the cost of the computation cost. It can be seen that PDLDS outperforms regular strategies on training time by 50\% on average.

\begin{table}[htbp]
  \centering
  \caption{Comparison results of defense success rates for FGSM and PGD attacks}
    \begin{tabular}[l]{cccc}
    \toprule
    Defense methods & FGSM & PGD \\
    \midrule
    Adversarial-PGD & 57.73\% & 55.72\% \\
    Adversarial-Network & 77.23\% & 75.04\% \\
    Adversarial-Dual-Network & 82.71\% & 80.92\% \\
    \textbf{Ours}  & 84.26\%(d=100) & 21.23\%(d=100) \\
    \bottomrule
    \end{tabular}%
  \label{tab16}%
\end{table}%

\section{conclusion}\label{sec:conclusion}

Deploying DLaaS in IoT faces security challenges. Furthermore, observing the impacts of fewer pixel mutations problem on the DNN, we developed a cooperative framework, PDLDS, which coordinates pixel mutations searching and filtering to decrease computation time for defense strategy. The basic idea is that not all pixels are involved in the filter defense approach. Only the pixels with slight change causing the DNN with the wrong prediction result will be filtered out. In this way, the computation time for the defense method can be effectively decreased without lowering the defense success rate. Experimental results showed 
PDLDS outperformed regular strategies on computation time by 50\% on average. In future work, we plan to implement in the real autonomous driving systems.

\bibliographystyle{IEEEtran}
\bibliography{references}

\end{document}